\documentclass[aip,jcp,reprint,floatfix]{revtex4-1}
\setcitestyle{super}
\usepackage{float}
\usepackage{caption}
\usepackage{graphicx}

\usepackage{amsfonts, amsmath, amssymb,latexsym}
\usepackage{bm}
\usepackage{dcolumn}
\usepackage[usenames, dvipsnames]{color}
\usepackage{bm}

\usepackage{subcaption}

\makeatletter
\renewcommand\@makecaption[2]{%
  \par
  \vskip\abovecaptionskip
  \begingroup
   \small\rmfamily
    \begingroup
     \samepage
     \flushing
     \let\footnote\@footnotemark@gobble
     \@make@capt@title{#1}{#2}\par
    \endgroup
  \endgroup
  \vskip\belowcaptionskip
}
\makeatother

\newcommand{\dif}{\mathrm{d}}

\begin{document}
\renewcommand{\thepage}{\arabic{page}}
\title{Shapes of Non-symmetric Capillary Bridges}

\author{L. R. Pratt} 
\affiliation{Department of Chemical and Biomolecular Engineering, Tulane University, New Orleans, LA 70118}

\author{D. T. Gomez} 
\affiliation{Department of Chemical and Biomolecular Engineering, 
Tulane University, New Orleans, LA 70118}

\author{A. Muralidharan} 
\affiliation{Department of Chemical and Biomolecular Engineering, 
Tulane University, New Orleans, LA 70118}

\author{N. Pesika} 
\affiliation{Department of Chemical and Biomolecular Engineering, 
Tulane University, New Orleans, LA 70118}

\begin{abstract}
Here we study the shapes of droplets captured between chemically
distinct parallel plates.  This work is a preliminary step toward
characterizing the influence of second-phase bridging between 
biomolecular surfaces on their solution contacts, \emph{i.e.,} capillary attraction or repulsion.   We
obtain a simple, variable-separated quadrature formula for the bridge
shape. The technical complication of double-ended boundary conditions on
the shapes of non-symmetric bridges is addressed by studying
\emph{waists} in the bridge shape, \emph{i.e.,} points where the bridge
silhouette has zero derivative.   Waists are always expected with symmetric
bridges, but waist-points can serve to characterize shape segments in
general cases. We study how waist possibilities depend on the physical
input to these problems, noting that these formulae change 
with the \emph{sign} of the inside-outside  pressure difference of the
bridge. These results permit a  variety of different interesting shapes,
and the development below is accompanied by several examples.
\end{abstract}

\date{\today}
\maketitle

\section{Introduction}

\begin{figure}[h]
\includegraphics[width=0.45\textwidth]{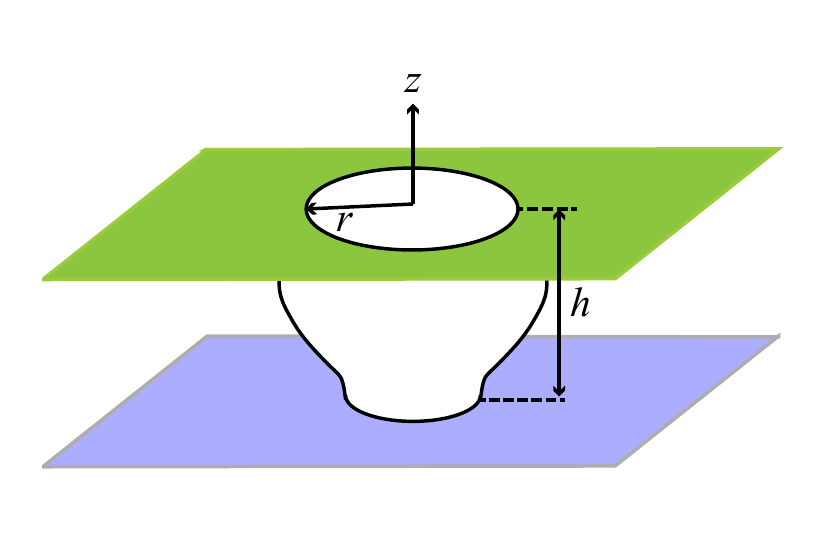}
\includegraphics[width=0.30\textwidth]{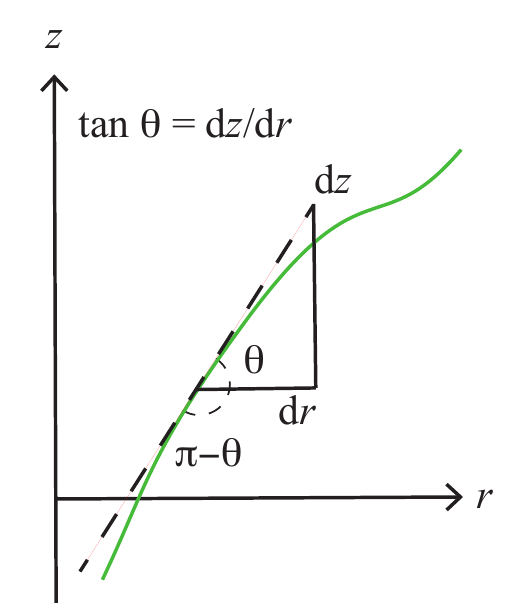}
\caption{(above) Non-symmetric capillary bridge studied here, and
(below) angles associated with a general droplet 
shape.}
  \label{fig:CapturedDrop}
\end{figure}

Here we study the shapes of non-symmetric capillary bridges between
planar contacts (FIG.~\ref{fig:CapturedDrop}), laying a basis for 
studying the forces that result from the bridging. 

The recent measurements of Cremaldi, \emph{et
al.,}\cite{Cremaldi:2015ft} provide a specific motivation for this work.
A helpful monograph\cite{de2013capillarity} sketches adhesion due to
symmetric capillary bridges, albeit with  aspect 
ratio (width/length $\approx$ 10$^3$) vastly different 
than is considered below. 
Additionally, that sketch\cite{de2013capillarity}  does not specifically consider non-symmetric
cases surveyed by Cremaldi, \emph{et al.}\cite{Cremaldi:2015ft} A
specific  description applicable to non-symmetric cases is apparently
unavailable,\cite{Lv:2018et} and, thus,  is warranted here.

A background aspect of our curiosity in these problems is the
possibility of evaporative bridging between ideal hydrophobic surfaces,
influencing the solution contacts between
biomolecules.\cite{10.1021/jp010945i,10.1073/pnas.1934837100,19605,10.1021/ja0441817,Cerdeirina:2011ix,dzubiella2006coupling,bharti2016capillary} 
Assessment of critical evaporative lengths in standard aqueous
circumstances on the basis of explicit 
thermophysical properties\cite{Cerdeirina:2011ix} sets
those lengths near 1~$\mu$m. Though we do not
specifically discuss that topic further here, our analytical development
does hinge on identification of the length $ \ell = 2\gamma/\vert \Delta
p\vert $, with $\gamma$ the fluid interfacial tension, and $\Delta p$
the pressure difference between inside and outside of the bridge.   The
experiments that motivate this study considered spans
$\lesssim\left(6\mu \mathrm{L}\right)^{1/3}\approx
1.8$~mm.\cite{Cremaldi:2015ft}

A full development of the essential basics of this problem might
be dense in statistical-thermodyamics.  We strive for
concision in the presentation below but 
follow\cite{de2013capillarity} a Grand Ensemble
formulation of our problem. We then develop the
optimization approach analogous to  Hamilton's 
Principle of 
classical mechanics.\cite{de2013capillarity,goldstein} That 
approach 
avoids more
subtle issues of differential geometry related to 
interfacial forces, and, eventually, should clarify the 
thermodynamic forces for 
displacement of the confining plates.  Along the way,  
we support 
the theoretical development by 
displaying  typical solutions of our formulation.

\section{Statistical Thermodynamic Formulation}
\begin{figure}[ht]
\begin{center}
\includegraphics[width=0.35\textwidth]{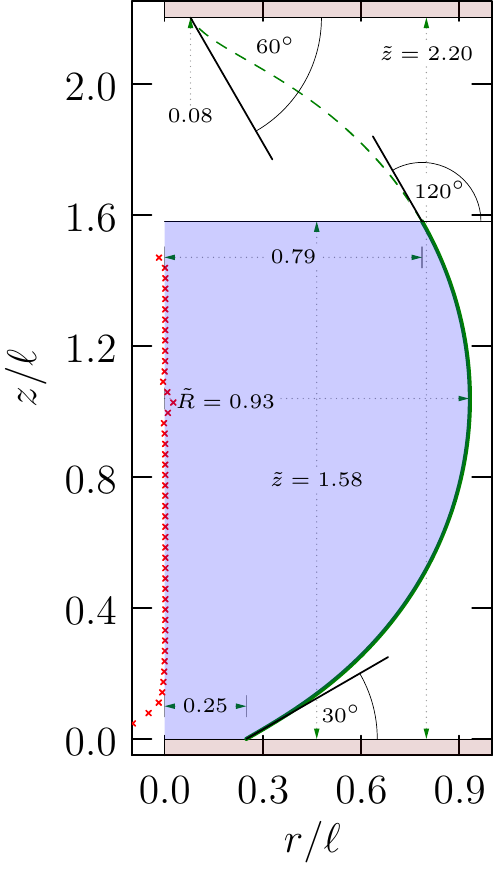}
\end{center}
\caption{Droplet dimensions using lengths scaled by $\ell = 2\gamma/\Delta  p$,
with $\Delta  p > 0$.  That the pressure  is higher inside than
outside the droplet is recognized by noting that $\ddot{r}$ is
negative at the waist. The separation of variables of
Eq.~\eqref{eq:scaled3again} suggests taking $r$ (the horizontal axis) as
the independent variable. At the bottom contact $\tilde{r}_-=
\frac{1}{2}\sin \theta_-$ in Eq.~\eqref{eq:waists0} with $\theta =
30^\circ$.  The waist has radius $\tilde{R} = \frac{1}{2}\left(1 + \cos
\theta_-\right) \approx 0.93$. The alternative solution of
Eq.~\eqref{eq:waists0} is $\frac{1}{2}\left(1 - \cos \theta_-\right)
\approx 0.067$, smaller than the radius of the upper contact, $ \approx
0.08$.  The  contact angle $\theta_+ = 60^\circ$ together with
$\tilde{R}$, Eq.~\eqref{eq:waists0} gives $\tilde{r}_+ \approx 0.79$,
confirming the connection between branches above and below the waist. The dashed curve thus
extends the solid curve. At each height, the red crosses  mark the
discrepancies of the Euler-Lagrange Eq.~\eqref{eq:optimize} from zero.}
\label{fig:CapBridgeProfile}
\end{figure}

Consider two plates, not necessarily the same, oriented perpendicular to
the $z$-axis and separated by a distance $h$ (FIG.
\ref{fig:CapturedDrop}). A droplet captured between two parallel plates
is assumed to be cylindrically symmetric about the $z$-axis.  We want to
determine the droplet shape (FIG. \ref{fig:CapBridgeProfile}) in advance
of analyses of the forces involved. We study 
\begin{multline}
\Delta \Omega\left\lbrack r\right\rbrack = - 2 \pi \Delta p \int\limits_{-h/2}^{h/2} \left(r^2/2\right) \mathrm{d}z \\
+ 2 \pi \gamma \int\limits_{-h/2}^{h/2}  r  \sqrt{1 + \dot{r}{}^2}\;\mathrm{d}z  
 \\
+  \pi r_+{}^2 \Delta \gamma_+ +  \pi r_-{}^2\Delta \gamma_-~,  
\label{eq:gcpotential}
    \end{multline}
a functional of the droplet radius $r(z)$. Here $\dot{r}
=\mathrm{d}r(z)/\mathrm{d} z $ and  $r_\pm = r(z= \pm h/2)$.  $\gamma$
is the tension between the droplet  and the external solution.  $\Delta
\gamma_+$ is the inside-outside difference of the surface tensions of
the fluids against the plate at $z=+ h/2$ (and similarly for $\Delta
\gamma_-$ with the fluids against the plate at $z=- h/2$); this
differencing will be clarified below as we note how this leads to
Young's Law.  $ \Delta p$ is the traditional Laplace
inside-outside pressure difference  of the bridge. The usual Grand
Ensemble potential for a single-phase uniform fluid solution being
$-\Omega = pV$, it is natural that $\Delta \Omega\left\lbrack
r\right\rbrack$ of Eq.~\eqref{eq:gcpotential} has $\Omega$ for
the surrounding fluid solution subtracted away; \emph{i.e}, the
pressure-volume term of Eq.~\eqref{eq:gcpotential} evaluates the
pressure-inside times the bridge volume, \emph{minus} the
pressure-outside times the same bridge volume.  Formally
\begin{eqnarray}
- \Omega = -F+ \sum_\alpha \mu_\alpha n_\alpha~,
\end{eqnarray}
with $F$ the Helmholtz free energy.  Therefore, the surface-area feature
of Eq.~\eqref{eq:gcpotential} can be viewed as an addition of $\gamma A$
contribution to $F$, with $A$ the surface area of contact of the bridge
with the external fluid and $\gamma$ is the tension of the fluid-fluid
interface.  

An alternative perspective on $\Delta \Omega$
(Eq.~\eqref{eq:gcpotential}) is that it is a Lagrangian function for
finding a minimum surface area of the bridge satisfying a given value of
the bridge volume.  Then $\Delta  p/ \gamma$, which has dimensions of an inverse length,
serves as a Lagrange multiplier.   We then minimize $\Delta \Omega$ with
respect to variations of $r(z)$, targeting a specific value of the
droplet volume.

The first-order variation of $\Delta\Omega$ is then
\begin{multline}
  \frac{\delta \Delta \Omega}{2 \pi} =
  -\Delta p \int\limits_{-h/2}^{h/2}  r \delta r \mathrm{d}z \\
    + \gamma\int\limits_{-h/2}^{h/2} \left\{ \left(\frac{ \dot{r}r}{\sqrt{{1 + \dot{r}^2}}}\right)\delta \dot{r}  + \left(\sqrt{1 + \dot{r}^2} \right)\delta r \right\} \mathrm{d}z \\
   +  r_+ \Delta \gamma_+\delta r_+  +  r_-\Delta \gamma_- \delta r_- ~.
\label{eq:1stvariationa}
\end{multline} 
The angle that the shape curve $r(z)$ makes with the plane perpendicular
to the $z$ axis (FIG.~\ref{fig:CapturedDrop}) is
\begin{eqnarray}
\cos^2 \theta = \frac{\dot{r}{}^2}{1 +\dot{r} {}^2 }~,
\end{eqnarray}
and at the contacting surfaces
\begin{eqnarray}
\left(\mp\right)\cos \theta_\pm = \frac{\dot{r}_\pm}{\sqrt{1 +\dot{r}_\pm {}^2 }}~.
\label{eq:contactangles}
\end{eqnarray}
Depicted in FIG.~\ref{fig:CapturedDrop} is the choice of the bottom sign
above, where $0<\dot{r}<\infty$.  For $\theta_+$
we change the choice so that the contact angle at the upper plate  is the traditional
external angle of the droplet.

The usual integration-by-parts for Eq.~\eqref{eq:1stvariationa} gives
\begin{multline}
 \frac{\delta \Omega}{2 \pi} = \\
\int\limits_{-h/2}^{h/2} \left 
\{- \gamma \frac{\mathrm{d}}{\mathrm{d}z}
\left ( \frac{ \dot{r}r}{\sqrt{1+\dot{r}^2}}\right) + \gamma \sqrt{1 + \dot{r}^2} 
- r \Delta p  \right \}\delta r\mathrm{d}z 
\\
+ \left(\Delta \gamma_+ + \frac{\gamma \dot{r}_+}{\sqrt{1 +\dot{r}_+ {}^2 }}\right)r_+ \delta r_+ \\
+ 
\left(\Delta \gamma_- -  \frac{\gamma \dot{r}_-}{\sqrt{1 +\dot{r}_- {}^2 }}\right)r_- \delta r_- ~. 
\label{eq:ibparts0}
\end{multline}

With the signs indicated in Eq.~\eqref{eq:contactangles} 
\begin{multline}
 \frac{\delta \Delta\Omega}{2 \pi} = \\
\int\limits_{-h/2}^{h/2} \left 
\{- \gamma \frac{\mathrm{d}}{\mathrm{d}z}
\left ( \frac{ \dot{r}r}{\sqrt{1+\dot{r}^2}}\right) + \gamma \sqrt{1 + \dot{r}^2} 
- r \Delta p  \right \}\delta r\mathrm{d}z 
\\
+ \left(\Delta \gamma_+ - \gamma \cos \theta_+\right)r_+ \delta r_+ \\
+ 
\left(\Delta \gamma_- - \gamma \cos \theta_-\right)r_- \delta r_- ~, 
\label{eq:ibparts1}
\end{multline}
with the exterior angles contacting the upper and lower plates.

The contact terms in Eq.~\eqref{eq:ibparts1} vanish if the contact
angles obey the  force balance 
\begin{eqnarray}
 \Delta \gamma_\pm = \gamma \cos \theta_\pm
 \label{eq:YoungsLaw}
\end{eqnarray}
of the traditional Young's Law.
This re-inforces the sign choice for Eq.~\eqref{eq:contactangles}.
Eq.~\eqref{eq:YoungsLaw} will provide boundary information for $r(z)$.  

From Eq.~\eqref{eq:ibparts1}, we require that the kernel  
\begin{eqnarray}
- \gamma \frac{\mathrm{d}}{\mathrm{d}z}
\left ( \frac{ \dot{r}r}{\sqrt{1+\dot{r}^2}}\right) + 
\gamma \sqrt{1 + \dot{r}^2} 
- r \Delta p  =0 ~
\label{eq:optimize}
\end{eqnarray}
vanish identically in $z$. As with Young's Law, this balances the forces
for  varying the droplet radius. For the example of a spherical droplet
of radius $R$, this force balance implies the traditional Laplace
pressure formula, $ \Delta p  = 2\gamma/R$.

The traditional Hamilton's principle\cite{goldstein} analysis of this
formulation then yields the usual energy conservation
theorem\cite{de2013capillarity,goldstein}
\begin{eqnarray}
\frac{\gamma  r }{\sqrt{1 + \dot{r}{}^2}}  - r^2 \Delta p /2 = D~,
\label{eq:conservationofenergy0}
\end{eqnarray}
with $D$ a constant of integration. $ D + r^2 \Delta p /2$ is non-negative
according to Eq.~\eqref{eq:conservationofenergy0}. Recognizing that
sign, then 
\begin{eqnarray}
\gamma  r \sin \theta\left(z\right)  = D+ r^2 \Delta p /2 ~,
\label{eq:conservationofenergy1}
\end{eqnarray}
with $0\le\theta\left(z\right) \le \pi$. The constant $D$ can be
eliminated in terms of boundary information, \emph{e.g.,}
\begin{eqnarray}
\gamma r_-\sin\theta_-  -  r_-^2 \Delta p /2 = D ~.
\end{eqnarray}
This helpfully correlates $r(z)$ at other places
too.  For example, we will consider
(FIG.~\ref{fig:CapBridgeProfile}) intermediate positions where $
\dot{r}\left(z\right)=0 $ and $\sin \theta\left(z\right)=1$. We call
such a position a \emph{`waist.'} A waist is expected for
symmetric cases that we build from here. Denoting the radius of a waist
by $R$, then
\begin{eqnarray}
 \gamma R = D + R^2 \Delta p /2~,
 \label{eq:positivehere}
\end{eqnarray}
from Eq.~\eqref{eq:conservationofenergy0}. This eliminates the
integration constant $D$ in favor of $R$ which may be more meaningful. 

\subsection{$\Delta  p > 0$}
\begin{figure}
\includegraphics[width=0.45\textwidth]{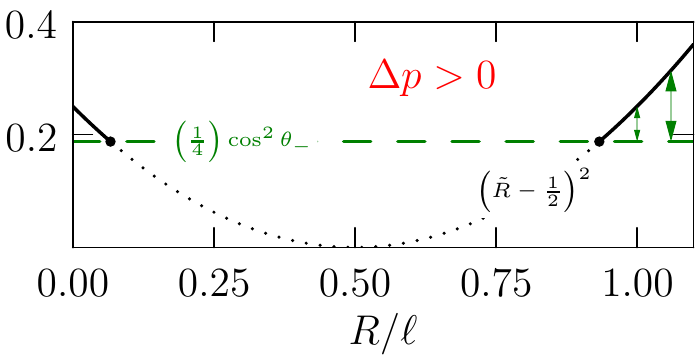}
\caption{For contact angle $\theta_-$, Eq.~\eqref{eq:waists0} requires
that $\left(\tilde{R}-1/2\right)^2 \ge \frac{1}{4} \cos^2\theta_-$. Thus,
the solid black curves cover possible values of $\tilde{R}$ for this
$\theta_-$, and displacements upward from the green horizontal line,
\emph{i.e.,} the arrows, show values of $\left(\tilde{r}_- -  \sin\theta_- 
/2 \right)^2$. The $\theta_-$ adopted for this drawing is $\pi/6$ as for
the bottom branch shown in FIG.~\ref{fig:CapBridgeProfile}, and the
right-most dot locates the value of the waist radius there
(FIG.~\ref{fig:CapBridgeProfile}).  Thus, the waist in that example is
the slimmest waist in that range.  Such considerations apply  to both
top and bottom contacts with their distinct contact angles.  A contact
angle near $\pi/2$ will correspond to a lower level for the horizontal line,
and thus be less restrictive of the possible values of a common waist
radius $\tilde{R}$.}
\label{fig:Rsizing}
\end{figure}

Considering $\Delta  p > 0$ we can make these relations more transparent by
non-dimensionalizing them with the length $ \ell = 2\gamma/ \Delta  p $.  Then 
$r = \tilde{r}\ell$ and $R = \tilde{R}\ell$, so 
\begin{eqnarray}
\tilde{r}_-\left(\tilde{r}_- -  \sin\theta_-\right) 
& = & \tilde{R}\left(\tilde{R}-1\right) ~. 
\label{eq:scaled5}
\end{eqnarray} 
Though this  scaling with the length $ \ell$ is algebraically
convenient, $ \Delta  p $ can take different signs in different settings; indeed
calculating from Eq.~\eqref{eq:optimize}, at a waist $ \Delta  p /\gamma =  1/R -
\ddot{r}$  in the present set-up, with $\ddot{r}$ the curvature at
that waist. Completing the square from Eq.~\eqref{eq:scaled5} gives
\begin{eqnarray}
\left(\tilde{R}-\frac{1}{2}\right)^2 =
\left(\tilde{r}_- -  \frac{ \sin\theta_-}{2} \right)^2 + \left(\frac{ \cos\theta_-}{2}\right)^2~. 
\label{eq:waists0}
\end{eqnarray}
Eq.~\eqref{eq:waists0} provides helpful perspective
(FIG.~\ref{fig:Rsizing}) for exploring different bridge sizes.  Given 
$\theta_- $, this requires that $\left(\tilde{R}-1/2\right)^2 \ge \left(
\cos\theta_- /2\right)^2$, as is evident there.

Interesting further consequences follow
from considerations of the cases that 
the droplet is nearly tangent
to the contact surfaces: $\theta_\pm = 0$ or $\pi$.  Consider first 
$\theta_- \rightarrow 0$. The 
droplet approaches detachment from the lower surface.  
We expect $r_- \rightarrow 0$ then.  FIG.~\ref{fig:Rsizing} shows that this 
can be achieved with $\tilde{R} = 0$ or 1.  The $\tilde{R} = 1$ case
produces a hemispherical lower portion on the bridge, with the hemisphere
just touching the lower surface and $\tilde{r}_- \approx 
(\frac{1}{2})\sin \theta_- $ from Eq.~\eqref{eq:waists0}.

When $\theta_+ \rightarrow \pi$ for 
example, the droplet preferentially wets the upper surface.  
We expect $r_+$ to be relatively large then, and
this force contribution describes inter-plate attraction, though 
not necessarily with a waist.

\subsection{More generally but $\Delta  p   \ne 0$}
Restoring in Eqs.~\eqref{eq:scaled5} and \eqref{eq:waists0} the
dependence on $\ell = 2\gamma/\Delta  p $ for  $\Delta  p \ne 0$, though possibly
negative, then gives
\begin{subequations}
\begin{eqnarray}
r_-\left(r_- -  \ell\sin\theta_-\right) 
& = & R\left(R-\ell\right), 
\label{eq:scaled5A}
\end{eqnarray}
\begin{multline}
\left(R - \frac{\ell}{2}\right)^2 =
\left(r_-  - \frac{ \ell\sin\theta_-}{2} \right)^2 \\
+ \left(\frac{ \ell\cos\theta_-}{2}\right)^2.
\label{eq:waists0B}
\end{multline}
\end{subequations}

$\ell$ is a \underline{\emph{signed}} length here. With these notations,
\begin{multline}
\cot^2 \theta = \frac{r^2\ell^2 - 
\left\lbrack r^2  - r_- \left(r_- - \ell \sin\theta_- \right) 
\right\rbrack^2
}{
\left\lbrack
r^2 - r_- \left( r_- - \ell \sin\theta_- \right) 
\right\rbrack^2
} \\
=  \left(\frac{\dif r}{\dif z}\right)^2
\label{eq:eliminateDalready}
\end{multline}
and
\begin{eqnarray}
\pm\dif z =  
\frac{
\left\lbrack
r^2 - r_- \left(r_- - \ell \sin\theta_- \right) 
\right\rbrack\dif r
}{
\sqrt{
r^2\ell^2 - 
\left\lbrack
r^2 - r_- \left(r_- - \ell \sin\theta_- \right) 
\right\rbrack^2
}
}
\label{eq:scaled3again}
\end{eqnarray}
separates these variables for integration.

\begin{figure}
\includegraphics[width=0.45\textwidth]{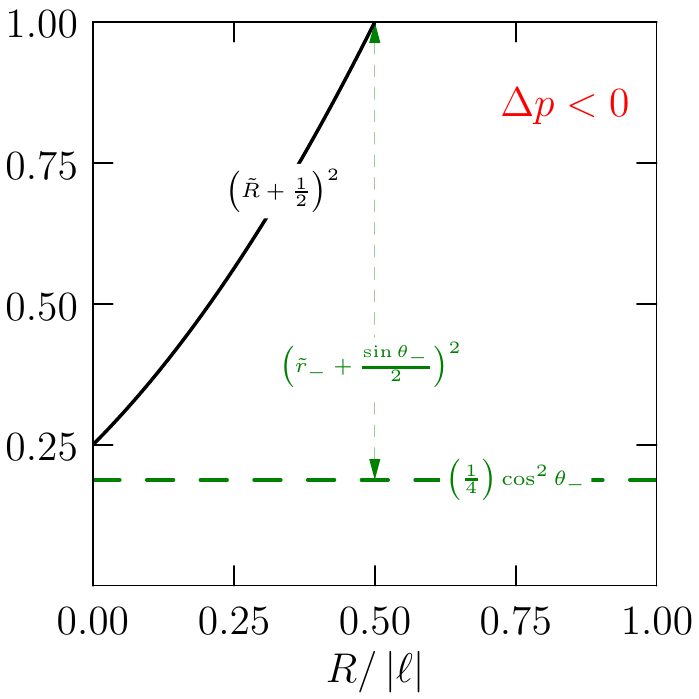}
\caption{Analog of FIG.~\ref{fig:Rsizing} but for the case $\Delta  p < 0$. See
Eq.~\eqref{eq:waists0CNeg}.}
  \label{fig:RsizingCNeg}
\end{figure}

We can still follow scaled lengths $\tilde{r} = r/\left\vert
\ell\right\vert$ and $\tilde{R} = R/\left\vert \ell\right\vert$.  Then
the analogue of Eq.~\eqref{eq:waists0} is 
\begin{eqnarray}
\left(\tilde{R}+\frac{1}{2}\right)^2 =
\left(\tilde{r}_- +  \frac{ \sin\theta_-}{2} \right)^2 + \left(\frac{ \cos\theta_-}{2}\right)^2~,
\label{eq:waists0CNeg}
\end{eqnarray}
when $\Delta  p < 0$; see FIG.~\ref{fig:RsizingCNeg}.  The analogue of
Eq.~\eqref{eq:scaled3again} with this length scaling for $\Delta  p < 0$  is
\begin{eqnarray}
\pm\dif \tilde{z} =  
\frac{
\left\lbrack
 \tilde{r}^2 -  \tilde{r}_- \left( \tilde{r}_- + \sin\theta_- \right) 
\right\rbrack\dif  \tilde{r}
}{
\sqrt{
 \tilde{r}^2 - 
\left\lbrack
 \tilde{r}^2 -  \tilde{r}_- \left( \tilde{r}_-  +  \sin\theta_- \right) 
\right\rbrack^2
}
}
\label{eq:scaled3againCNeg}
\end{eqnarray}
To achieve $\Delta  p / \gamma =  1/R - \ddot{r} < 0$ for a bridge with 
wiast radius
$R$, clearly the curvature $\ddot{r}$ at that waist should be
substantially positive to ensure that the negative second contribution dominates. 
In addition, the radius at the waist should be fairly large, thereby
reducing the contribution of the positive first term.  These 
points combined suggest
that to achieve adhesion the contact areas should be larger than the waist area, 
which itself should be substantial.

\subsection{Waist $R$}
Reaffirming the identification of  $R$ as the radius of a waist, and specifically 
recalling that
$\ell$ is a \underline{\emph{signed}} length:
\begin{multline}
\cot^2 \theta\left(z\right)  =  \frac{
r^2\ell^2 - \left\lbrack \ell R + \left(r^2- R^2\right)\right\rbrack^2
}{
\left\lbrack \ell R + \left( r^2- R^2\right)\right\rbrack^2
}  =  \left(\frac{\dif r}{\dif z}\right)^2~.
\label{eq:scaled0}
\end{multline}
Factoring-out the  $\cot^2 \theta
\left(\tilde{r}^2=\tilde{R}^2\right)=0$ feature gives
\begin{multline}
\cot^2 \theta\left(z\right) =  
\frac{
\left(R^2 - r^2\right)\left(r^2 - (R-\ell)^2 \right)
}{
\left\lbrack r^2 - R \left(R-\ell\right)\right\rbrack^2
} =  \left(\frac{\dif r}{\dif z} \right)^2~.
\label{eq:scaled2}
\end{multline}
Eq.~\eqref{eq:scaled2} also shows that $\cot^2 \theta\left( z\right) = 0$ at the point
$ r^2 =  (R-\ell)^2$.

Eq.~\eqref{eq:scaled2} then achieves the  separation of variables 
\begin{eqnarray}
\pm\dif z  =  
\frac{
\left\lbrack r^2 -  R \left( R-\ell \right)\right\rbrack\dif  r
}{
\sqrt{
\left( R^2 - r ^2\right)\left( r^2 - \left( R-\ell \right)^2 \right)
}
}
\label{eq:scaled3}
\end{eqnarray}
for integration in this case.

\section{Examples}
In the example FIG.~\ref{fig:CapBridgeProfile} ($\Delta  p > 0$),
$\tilde{R}\approx 0.933$ and $\left( \tilde{R}-1\right)^2 \approx
0.067{\,}^2$, smaller than the radius of the upper cross-section,
$0.08{\,}^2$, in that extended example. The slimmer \emph{second
waist} is not realized.

\begin{figure}
\includegraphics[width=0.45\textwidth]{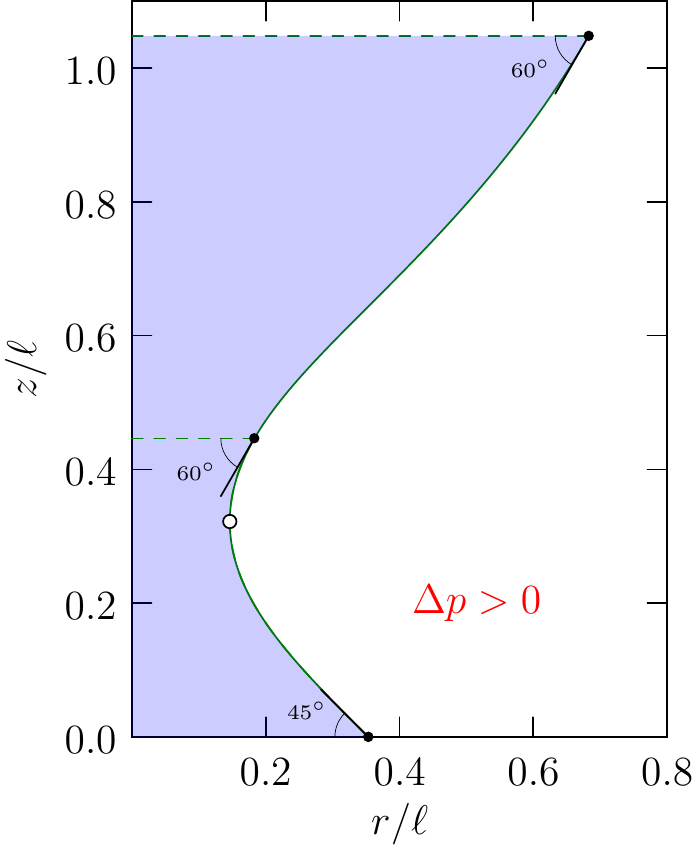}
\caption{Capillary bridge shape for the biggest slim-waisted 
possibility of
FIG.~\ref{fig:RsizingSkinny}.  Here the pressure inside is greater
than the pressure outside.  The open circle marks the waist.  $\Delta  p / \gamma =  1/R - \ddot{r}$
at a 
waist of radius $R$, so achieving $ \Delta  p > 0 $ with positive curvature 
$\ddot{r}$, as above, limits the waist radius $R$. 
This
solution  exhibits the upper contact angle twice. }
\label{fig:CapBridgeProfileCompareSkinny}
\end{figure}

\begin{figure}
\includegraphics[width=0.45\textwidth]{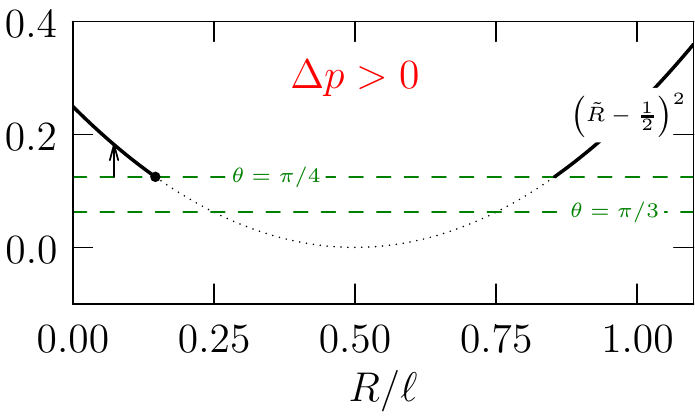}
\caption{Considerations for choice of waist radii $\tilde{R}$ for the slim-waisted
bridge of FIG.~\ref{fig:CapBridgeProfileCompareSkinny}.}
\label{fig:RsizingSkinny}
\end{figure}

FIG.~\ref{fig:CapBridgeProfileCompareSkinny} shows a bridge shape 
for the slender waist identified for the contact angles 
specified  in FIG.~\ref{fig:RsizingSkinny}
for $\Delta  p > 0$.

\begin{figure}
\includegraphics[width=0.45\textwidth]{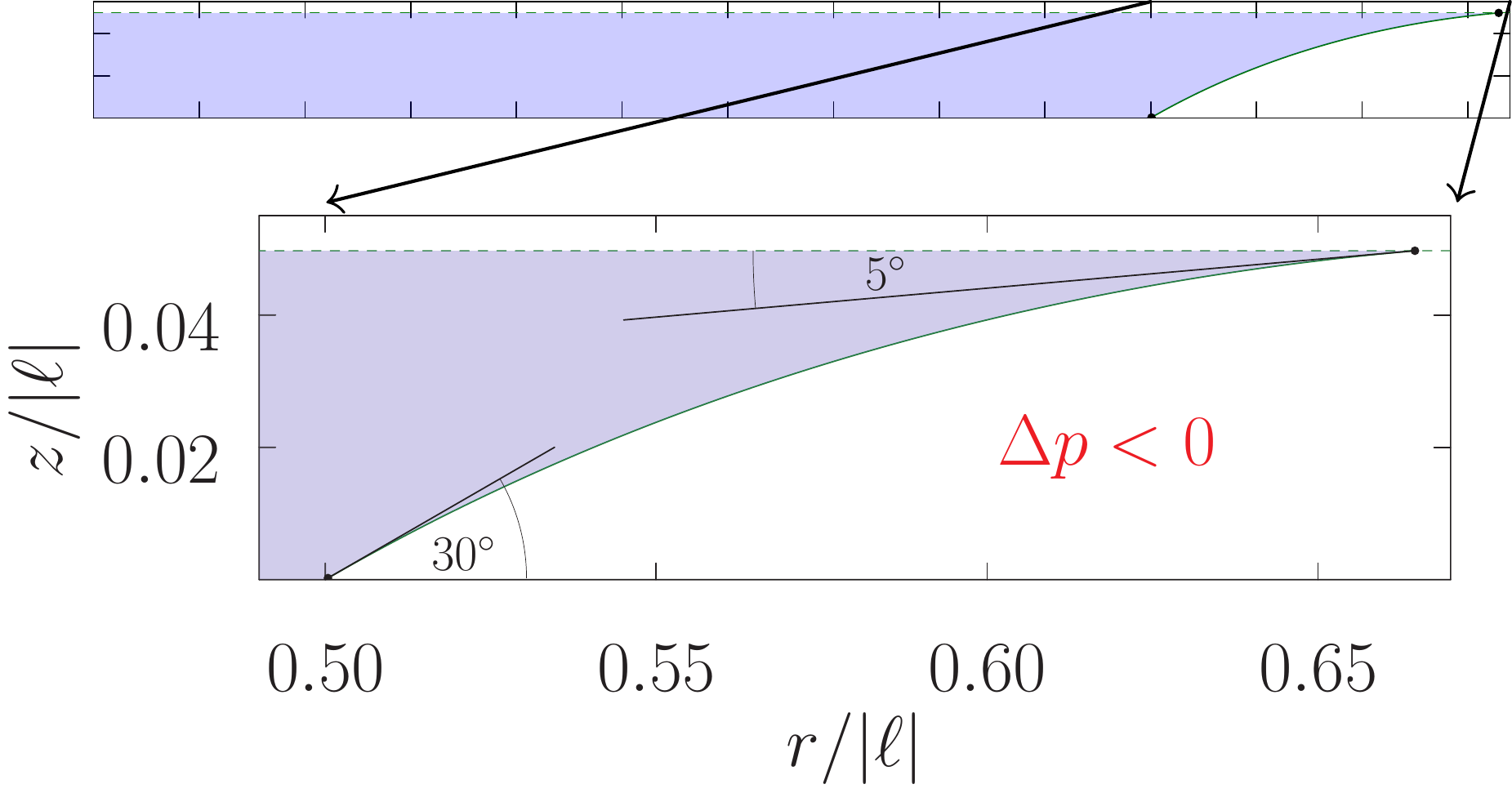}
\caption{From Eq.~\eqref{eq:scaled3}, with the indicated contact angles
and with $\Delta  p < 0$, so that pressure inside the bridge is less
than the pressure outside. $\tilde{R} \approx 0.366$, from
Eq.~\eqref{eq:waists0CNeg} and FIG.~\ref{fig:RsizingCNeg}.  Since the 
smallest contact radius --- at the bottom plate ---  is 0.5, the waist
at $\tilde{R} \approx 0.366$ is not realized in this physical range.  }
  \label{fig:CBCNegComposite}
\end{figure}

In the example FIG.~\ref{fig:CBCNegComposite}, $\Delta  p < 0$ and $\ell < 0$.
Thus $ \left( R-\ell \right) = \vert \ell \vert \left( \tilde{R}+1
\right)$, and the $(-)$ of Eq.~\eqref{eq:scaled3} is required to achieve
a positive slope at the bottom plate. The aspect ratio of the bridge 
is
vastly changed, as was true also in the discussion of capillary 
adhesion of Ref.~\citenum{de2013capillarity};  capillary adhesion would 
be expected for this shape.

\section{Discussion}
In view of the  variety of interesting shape possibilities, we reserve
explicit study of the consequent inter-plate forces for a specific
experimental context.  Nevertheless, we outline here how such a
practical study might be implemented.

The setup above permits straightforward calculation of the thermodynamic
potential $\Omega$, and 
\begin{eqnarray}
\frac{\dif \Omega}{\dif h} =  \left\langle \frac{\dif  U}{\dif h} \right\rangle = -F_h~. 
     \end{eqnarray}
$U$ being the internal energy, positive values of $F_h$ indicates that
$U$ decreases with increasing $h$, temperature being constant in these
considerations.  Thus, positive values of $F_h$ indicate repulsion, and
negative values describe attraction.

Our motivating example is Cremaldi, \emph{et
al.;}\cite{Cremaldi:2015ft} in those cases a waist with radius
$\tilde{R}$ is clear, and we anticipate that $\Delta  p > 0$.
To connect to specific experimental cases, we note that \emph{a priori}
experimental data are $\gamma$, the contact angles $\theta_-$
and $\theta_+$, the experimental volume of the captured droplet $v$, 
and inter-plate separation $h$. 
Eq.~\eqref{eq:waists0}  and FIG.~\ref{fig:Rsizing} show permitted ranges
for $\tilde{R}$.  With these parameters set, integration
(Eq.~\eqref{eq:scaled3}) determines $\Delta \tilde{z} = \tilde{z}_+ -
\tilde{z}_-$. Then 
\begin{eqnarray}
h = 
\left\vert \ell\right\vert \Delta \tilde{z}~,
\label{eq:ez0}
\end{eqnarray}
so that
\begin{eqnarray}
     \frac{1}{\left\vert \ell\right\vert} = \frac{\left\vert \Delta  p \right\vert}{2\gamma} = \frac{\Delta \tilde{z}}{h}~,
    \label{eq:ez1}
\end{eqnarray}
matching the experimental $h$.  [What is more, the \emph{\underline{sign}} of $\Delta p$ is known
through the calculational procedure.] 
We then further  evaluate the volume of droplet 
\begin{eqnarray}
    v\left\lbrack\tilde{R}\right\rbrack = 
   \left\vert  \ell\right\vert^3 \pi \int\limits_{\tilde{z}_-}^{\tilde{z}_+} 
    \tilde{r}^2\left(\tilde{z}\right) \dif \tilde{z}~
    \label{eq:v}
\end{eqnarray}
as it depends on $\tilde{R}$, and seek a match with the experimental
droplet volume $v$.  If $\tilde{R}$ were provided \emph{a priori},
Eqs.~\eqref{eq:ez1} and \eqref{eq:v} would over-determine $\ell$.   But
$\tilde{R}$ is not provided \emph{a priori}, so those two equations
determine the two remaining parameters $\ell$ and $\tilde{R}$.   Since
the dependence on $\left\vert \ell\right\vert$ is clear, we can proceed
further to 
\begin{eqnarray}
    v\left\lbrack\tilde{R}\right\rbrack = 
    \left(\frac{h}{\Delta \tilde{z}}\right)^3 
    \pi \int\limits_{\tilde{z}_-}^{\tilde{z}_+} 
    \tilde{r}^2\left(\tilde{z}\right) \dif \tilde{z}~,
    \label{eq:v1next}
\end{eqnarray}
leaving finally
\begin{eqnarray}
   \pi
    \int\limits_{\tilde{z}_-}^{\tilde{z}_+} 
    \tilde{r}^2\left(\tilde{z}\right) \dif \tilde{z}/\Delta \tilde{z}\;{}^3
    = \frac{v_0}{h^3}
    \label{eq:v2next}
\end{eqnarray}
to be solved for $\tilde{R}$. 

\section{Conclusions}
We provide general, simple, variable-separated quadrature formulae
(Eq.~\eqref{eq:scaled3}) for the  shapes  of  capillary bridges, not
necessarily symmetric. The technical complications of double-ended
boundary conditions on the shapes of non-symmetric bridges are
addressed by studying \emph{waists} in the bridge shapes,  noting that
these relations change distinctively with change-of-sign of the
inside-outside  pressure difference of the bridge
(Eq.~\eqref{eq:waists0B}). These results permit a variety
of different interesting cases, and we discuss how these analyses should be
implemented to study forces resulting from capillary bridging 
between neighboring surfaces in  solutions.


\providecommand{\latin}[1]{#1}
\makeatletter
\providecommand{\doi}
  {\begingroup\let\do\@makeother\dospecials
  \catcode`\{=1 \catcode`\}=2 \doi@aux}
\providecommand{\doi@aux}[1]{\endgroup\texttt{#1}}
\makeatother
\providecommand*\mcitethebibliography{\thebibliography}
\csname @ifundefined\endcsname{endmcitethebibliography}
  {\let\endmcitethebibliography\endthebibliography}{}

\end{document}